\documentclass[twocolumn,epjc3]{svjour3}

\usepackage{grffile}
\usepackage{slashed}
\usepackage{bbm}
\usepackage{footmisc}
\usepackage{multirow}

\usepackage{amssymb}
\usepackage{amsmath}
\usepackage{color}
\usepackage{xspace}
\usepackage{caption}
\usepackage{subcaption}

\usepackage{datetime}
\usepackage{color,fancybox}

\RequirePackage[T1]{fontenc}

\smartqed  

\RequirePackage{graphicx}
\RequirePackage{mathptmx}      
\RequirePackage{flushend}
\RequirePackage[numbers,sort&compress]{natbib}
\RequirePackage[colorlinks,citecolor=blue,urlcolor=blue,linkcolor=blue]{hyperref}

\journalname{Eur. Phys. J. C}

\newcommand{\GeV}{\ensuremath{\,\mathrm{GeV}}\xspace}

\newcommand{\fb}{\ensuremath{\,\mathrm{fb}}\xspace}

\newcommand{\order}[1]{\mathcal{O}\!\left(#1\right)}

\newcommand{\bea}{\begin{eqnarray}}
\newcommand{\eea}{\end{eqnarray}}

\newcommand{\crn}{\nonumber \\}
\newcommand{\kfac}{$K$-factor\xspace}

\newcommand{\eq}[1]{Eq.~(\ref{#1})}
\newcommand{\bib}[1]{Ref.~\cite{#1}}
\newcommand{\fig}[1]{Fig.~\ref{#1}}
\newcommand{\tab}[1]{Table~\ref{#1}}

\begin{document}
\title{Next-to-leading order QCD corrections to $W\gamma$ production
  in association with two jets}

\author{Francisco~Campanario \thanksref{e1,addr1} 
         \and
         Matthias~Kerner   \thanksref{e2,addr2} 
         \and
         Le~Duc~Ninh      \thanksref{e3,addr2,addr3} 
         \and
         Dieter~Zeppenfeld   \thanksref{e4,addr2} 
         }

\thankstext{e1}{e-mail: francisco.campanario@ific.uv.es}
\thankstext{e2}{e-mail: matthias.kerner@kit.edu}
\thankstext{e3}{e-mail: duc.le@kit.edu}
\thankstext{e4}{e-mail: dieter.zeppenfeld@kit.edu}

\institute{Theory Division, IFIC, University of Valencia-CSIC, E-46980
  Paterna, Valencia, Spain \label{addr1}
   \and 
    Institute for Theoretical Physics, KIT, 76128 Karlsruhe,
    Germany \label{addr2}
   \and 
   Institute of Physics, Vietnam Academy of Science and Technology, 10 Dao Tan, Ba Dinh, Hanoi, Vietnam \label{addr3}
   }

\journalname{FTUV-14-3001\;\;IFC/14-10\;\;KA-TP-02-2014\;\;LPN14-012\;\;SFB/CPP-14-10}


\maketitle

\begin{abstract}
The QCD-induced $W^\pm \gamma$ production channels in association with
two jets are computed at next-to-leading order QCD accuracy. The W bosons decay 
leptonicly and full off-shell and finite width effects as
well as spin correlations are taken into account. These processes are
important backgrounds to beyond Standard Model physics searches and also
relevant to test the nature of the quartic gauge couplings of the
Standard Model. The next-to-leading order corrections reduce the scale
uncertainty significantly and show a non-trivial phase space
dependence. Our code will be publicly available as part of the parton
level Monte Carlo program {\texttt{VBFNLO}}.
\end{abstract}

\section{Introduction}
Di-boson production in association with two jets constitutes an
important set of processes at the LHC. They are backgrounds to many
Standard Model (SM) searches. For example, W-, Z- and photon-pair
production with two accompanying jets are irreducible backgrounds of
Higgs production via vector boson fusion. Furthermore, they are sensitive to
triple and quartic gauge couplings, thereby providing us with an excellent avenue to
understand the electroweak (EW) sector of the SM and possibly to get hints of physics beyond the SM.

There are two mechanisms to produce them, namely, 
EW-induced channels of order $\order{\alpha^4}$ and QCD-induced processes of
order $\order{\alpha_s^2 \alpha^2}$ for on-shell production at leading order (LO). 
Additionally, the EW mode is classified into ``vector boson fusion'' (VBF) mechanism, which
involves $t$ and $u$ channel exchange, and $s$ channel contributions corresponding
mainly to $VVV$ production with 
one $V$ decaying into two jets. 

The VBF production modes include vector boson scattering, $VV\to
VV$, as a basic topology. For massive gauge boson scattering, the 
main interest will be to elucidate whether the recently discovered
Higgs boson unitarizes this process as predicted in the SM. 
Processes with a real photon in 
the final state are also interesting since they are sensitive to
triple and quartic gauge couplings and have a higher cross section.

The next-to-leading order (NLO) QCD corrections to the VBF processes have been computed in
Refs.~\cite{Jager:2006zc,Jager:2006cp,Bozzi:2007ur,Jager:2009xx,Denner:2012dz}
for all combinations of massive gauge bosons, including leptonic
decays of the gauge bosons as well as all off-shell and finite width effects. 
A similar calculation with a $W$ boson and a real photon in the final
state has been done in \bib{Campanario:2013eta}. For the $s$ channel
contributions, the NLO QCD corrections with leptonic decays were computed in
Refs.~\cite{Hankele:2007sb,Campanario:2008yg,Bozzi:2009ig,Bozzi:2010sj,Bozzi:2011en,Bozzi:2011wwa}
and are available via the {\texttt{VBFNLO}}
program~\cite{Arnold:2008rz,Arnold:2012xn} (see also
Refs.~\cite{Lazopoulos:2007ix,Binoth:2008kt,Baur:2010zf} for on-shell
production and \bib{Nhung:2013jta} for NLO EW corrections).

NLO QCD corrections to the QCD-induced processes have been computed
for
$W^+W^+jj$~\cite{Melia:2010bm,Melia:2011gk,Jager:2011ms,Campanario:2013gea},
$W^+W^-jj$~\cite{Melia:2011dw,Greiner:2012im}, $W^\pm
Zjj$~\cite{Campanario:2013qba} and $\gamma \gamma
jj$~\cite{Gehrmann:2013bga} production. 
Results for $\gamma \gamma jjj$ production at NLO QCD have been very recently presented in Ref.~\cite{Badger:2013ava}.

In this paper, we provide first results for the QCD-induced $W\gamma jj$
production channel. The calculation is based on our previous implementation of 
NLO QCD corrections to $WZjj$ production processes~\cite{Campanario:2013gea}, where the off-shell photon contribution was 
included. 
The interference effects between the QCD and EW induced amplitudes are
generally small for most
applications~\cite{Jager:2011ms,Denner:2012dz,Campanario:2013gea} and
are not considered here.
Leptonic decays of the $W$ boson as well as all off-shell
effects are consistently taken into account. This includes also the
radiative decay of the $W$ with a real photon radiated off a charged lepton, 
which diminishes the sensitivity of the 
EW-induced $W\gamma jj$ production mode to anomalous couplings. In this paper, 
we follow the approach of \bib{Baur:1993ir} (see also references therein) to reduce this contribution 
by imposing a cut on the transverse mass of the $W\gamma$ system. 

To define the $W\gamma jj$ signature, since our study is done at the jet cross section level and 
fragmentation contributions are not taken into account, the real photon has to be isolated from the partons to avoid 
collinear singularities due to $q \to q \gamma$ splittings. 
While a similar issue with the charged 
lepton can be resolved by imposing a simple cut on 
$R_{l\gamma}=[(y_\gamma - y_l)^2 + (\phi_\gamma - \phi_l)^2]^{1/2}$ ($y$ and $\phi$ being the 
the rapidity and azimuthal angle, respectively) to separate the photon from the charged lepton, it 
cannot be applied to partons because doing so would also remove events with a soft gluon. 
These events are needed at NLO (or beyond) to cancel soft divergences in the virtual amplitudes. 
To solve this problem, we use the smooth cone isolation cut proposed by Frixione~\cite{Frixione:1998jh}. 
This approach preserves IR safety without the use of fragmentation functions and thereby allows us to focus on the 
physics of the hard photon. 

The QCD-induced $W\gamma jj$ production process has been implemented within the {\texttt{VBFNLO}}
framework, a parton level Monte Carlo program
which allows the definition of general acceptance cuts and
distributions.

This paper is organized as follows: In the next section, the major
points of our implementation will be provided. In
Section~\ref{sec:results} the setup used for the calculation and the
numerical results for inclusive cross sections and various distributions will be given. 
Conclusions are presented in Section~\ref{sec:con} and in the
appendix results at the amplitude squared level for a random phase-space point are provided.
\section{Calculational details}
\label{sec:cal_details}
In this paper, we compute the QCD-induced processes at NLO QCD for
the process
\begin{equation}
\label{eq:process}
pp \to
l^{\pm}\overset{\textbf{\fontsize{0.5pt}{0.5pt}\selectfont(---)}}{\nu_{l}} \gamma\; jj + X,
\end{equation}
at order ${\cal O}(\alpha_s^3 \alpha^3)$. We present results for the specific leptonic final state
$e^{\pm}\overset{\textbf{\fontsize{0.5pt}{0.5pt}\selectfont(---)}}{\nu_{e}}\gamma$
and refer to the process as $W\gamma jj$ production for
simplicity. The final results can be multiplied by a factor two to
take the
$\mu^{\pm}\overset{\textbf{\fontsize{0.5pt}{0.5pt}\selectfont(---)}}{\nu_{\mu}}\gamma$
channels into account. To compute the amplitudes, we follow
the method described in Ref.~\cite{Campanario:2013qba} for $W^\pm Z jj$ production
implemented in the {\texttt{VBFNLO}} program. We provide a summary here for the sake of being self contained. 

The Feynman
diagrammatic approach is taken and for simplicity we choose to
describe the resonating $W^\pm$ propagators with a fixed width and
keep the weak-mixing angle real. At LO, we classify all contributions
into $4$-quark and $2$-quark-$2$-gluon amplitudes, e.g. for $W^+\gamma jj$
\begin{align}
u\bar{d} &\to \bar{u}u\; l^{+}\nu_{l} \gamma,\crn
u\bar{d} &\to \bar{c}c\; l^{+}\nu_{l} \gamma,\crn
u\bar{d} &\to \bar{d}d\; l^{+}\nu_{l} \gamma,\crn
u\bar{d} &\to \bar{s}s\; l^{+}\nu_{l} \gamma,\crn
gg &\to \bar{u}d\; l^{+}\nu_{l} \gamma 
\label{eq:subproc}
\end{align}
and accordingly for $W^-\gamma jj$.

\begin{figure}[h]
  \centering
\includegraphics[width=0.65\columnwidth]{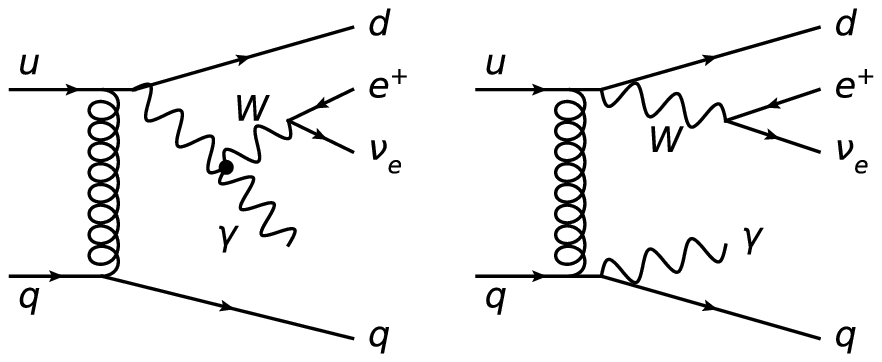}
\includegraphics[width=0.65\columnwidth]{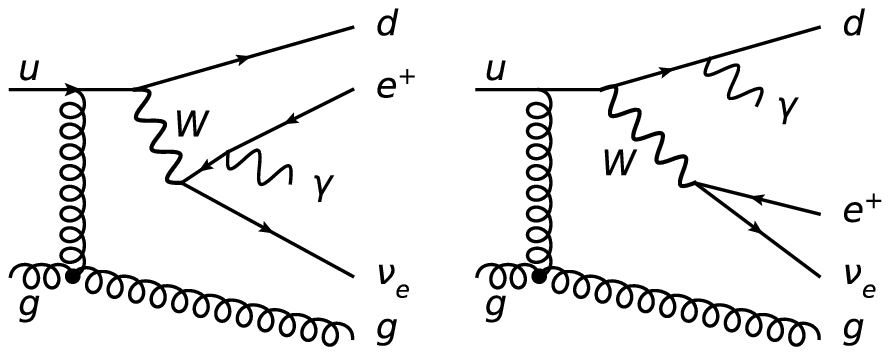}
\caption{Representative tree-level Feynman diagrams.}
\label{fig:feynTree}
\end{figure}

From these five generic subprocesses we can obtain all the amplitudes of other subprocesses via crossing. 
Some representative Feynman diagrams are displayed in Fig.~\ref{fig:feynTree}. 
We work in the 5-flavor scheme, hence the bottom-quark contribution with $m_b = 0$ is included. 
Subprocesses with external top quarks should be treated as different signatures and therefore are omitted. 
However, the virtual top-loop contribution is included in our calculation. 

At NLO QCD, there are the virtual and the real 
corrections. We use dimensional regularization~\cite{'tHooft:1972fi}
to regularize the ultraviolet (UV) and infrared (IR) divergences and
use an anticommuting prescription of
$\gamma_5$~\cite{Chanowitz:1979zu}. The UV divergences of the virtual
amplitude are removed by the renormalization of $\alpha_s$. Both the
virtual and the real corrections are infrared divergent. These
divergences are canceled using the Catani-Seymour prescription~\cite{Catani:1996vz} such
that the virtual and real corrections become separately numerically
integrable. As mentioned in the introduction, 
collinear singularities that result from a real
photon emitted off a massless quark are eliminated using the
{\it photon isolation cut} proposed by
Frixione, which preserves the IR QCD
cancellation and eliminates the need of introducing 
photon fragmentation functions. The real emission contribution
includes, allowing for external bottom quarks, $186$ subprocesses with six
particles in the final state. 
\begin{figure}[th]
  \centering
  \begin{subfigure}[b]{0.27\columnwidth}
    \centering
    \includegraphics[scale=0.65]{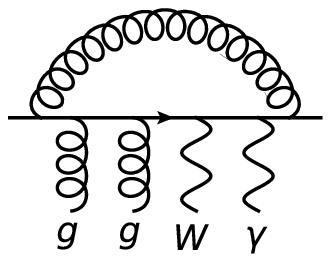}
    \caption{}
    \label{fig:HexLine}
  \end{subfigure} 
  \begin{subfigure}[b]{0.27\columnwidth}
    \centering
    \includegraphics[scale=0.65]{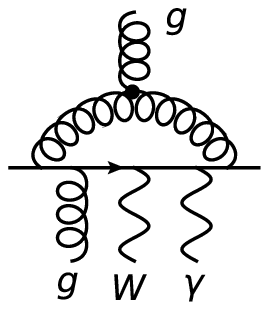}
   \caption{}
  \end{subfigure} 
  \begin{subfigure}[b]{0.27\columnwidth}
    \centering
    \includegraphics[scale=0.65]{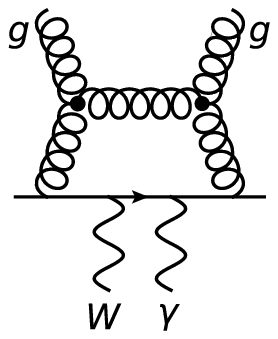}
    \caption{}
  \end{subfigure} 
  \begin{subfigure}[b]{0.27\columnwidth}
    \centering
    \includegraphics[scale=0.65]{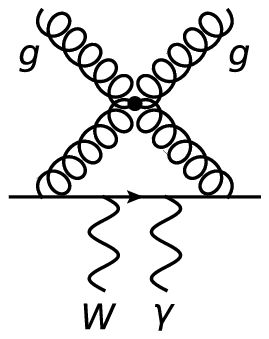}
   \caption{}
  \end{subfigure} 
  \begin{subfigure}[b]{0.27\columnwidth}
    \centering
    \includegraphics[scale=0.65]{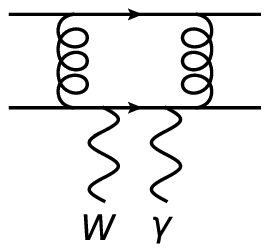}
    \caption{}
  \end{subfigure} 
  \begin{subfigure}[b]{0.27\columnwidth}
    \centering
    \includegraphics[scale=0.65]{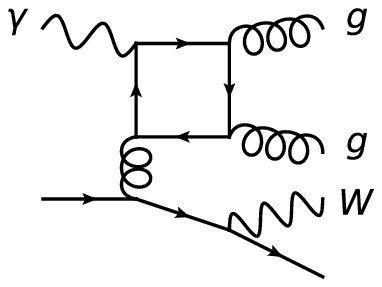}
   \caption{}
  \end{subfigure}
  \caption{Selected Feynman diagrams contributing to the virtual amplitudes.}
  \label{fig:feynVirt}
\end{figure}

The virtual amplitudes are more
challenging involving up to six-point rank-five one-loop
tensor integrals appearing in the $2$-quark-$2$-gluon virtual
amplitudes. There are $42$ six-point diagrams for
each of seven independent subprocesses. 
The $4$-quark group is simpler with only $6$ hexagons for
the most complicated subprocesses with same-generation quarks.
\fig{fig:feynVirt} shows some selected contributions to the virtual
amplitude. The evaluation of scalar integrals is done following 
Refs.~\cite{'tHooft:1978xw, Bern:1993kr,Dittmaier:2003bc,Nhung:2009pm,Denner:2010tr}. 
The tensor coefficients of the loop 
integrals are computed using the Passarino-Veltman reduction
formalism~\cite{Passarino:1978jh} up to the box level. 
For pentagons and hexagons, 
we use the reduction formalism of Ref.~\cite{Denner:2005nn} (see also
Refs.~\cite{Campanario:2011cs,Binoth:2005ff}).

Our calculation has been carefully checked as follows. 
The present code is adapted from our previous implementation of 
the $WVjj$ ($V=Z,\gamma^*$) production processes~\cite{Campanario:2013gea}, which has been 
crosschecked at the amplitude level by two independent calculations. 
The adaptation includes removing the $Z$ contribution, disallowing the decay $\gamma^* \to l^+l^-$ and 
adding the radiative decay $W^\pm \to l^{\pm}\overset{\textbf{\fontsize{0.5pt}{0.5pt}\selectfont(---)}}{\nu_{l}} \gamma$. 
These trivial changes are universal and have been crosschecked. 
Moreover, the real emission contributions have been crosschecked
against Sherpa~\cite{Gleisberg:2008ta,Gleisberg:2008fv} and agreement at the per mill level was found. 
A nontrivial change arises in the virtual amplitudes where we have to calculate a new set of scalar 
integrals which do not occur in the off-shell photon case. We have again checked this with two independent 
calculations and obtained full agreement at the amplitude level. 
The first implementation uses
{\texttt{FeynArts-3.4}}~\cite{Hahn:2000kx} and
{\texttt{FormCalc-6.2}} \cite{Hahn:1998yk} to obtain the virtual
amplitudes. The in-house library {\texttt{LoopInts}} is used to
evaluate the scalar and tensor one-loop integrals. 

In the following, we sketch the second implementation, which will be
publicly available via the {\texttt{VBFNLO}} program and is the one 
used to obtain the numerical results presented in the next section. 
As customary in all {\texttt{VBFNLO}} calculations, the
spinor-helicity formalism of~\bib{Hagiwara:1988pp} is used throughout 
the code.
The leptonic decays of the EW gauge bosons, which are common for all
sub-processes, are calculated once for each phase-space point and stored.
In addition we pre-calculate parts of Feynman diagrams, that are
common to the sub-processes of the real emission and use a
caching system to compute Born amplitudes appearing in different dipole
terms~\cite{Catani:1996vz} only once.

For the virtual amplitudes, we use generic building blocks, computed with the
in-house program described in Ref.~\cite{Campanario:2011cs}, which include
groups of loop corrections to Born topologies with a fixed number and
a fixed order of external particles, i.e. all self-energy, triangle,
box, pentagon and hexagon corrections to a quark line with four attached 
gauge bosons are combined into a single routine. The scalar and tensor
integrals are computed as described in Ref.~\cite{Campanario:2011cs}.

The control of the numerical instabilities is done as customary in our
calculations using Ward identities. By replacing a polarization vector
with the corresponding momentum, one can build up identities relating
$N$-point integrals to lower point integrals. This property is
transferred to the building blocks as described in
Ref.~\cite{Campanario:2011cs}, providing an additional check of the
correctness on the calculation of the virtual amplitudes. This
procedure is possible because we factorize the color and EW
couplings from the building blocks and assume the polarization vector
of the external gauge bosons as an effective current without using
special properties like transversality or on-shellness. 
These identities are called gauge tests and are checked for every
phase space point with a small additional computing cost by using a
cache system. 
If the gauge tests are true by less than $2$ digits with double
precision, the program recalculates the associated building blocks
with quadruple precision and the point is discarded if the gauge
tests still fail. 
After this step, the number of discarded points is statistically
negligible for a typical calculation with the inclusive cuts specified
in the next section. 
This strategy was also successfully applied in, e.g.,
Refs.~\cite{Campanario:2011ud,Campanario:2012bh,Campanario:2013qba,Campanario:2013mga,Campanario:2013gea}.
With this method, we obtain the NLO inclusive cross section with
statistical error of $1\%$ in three hours on an Intel $i5$-$3470$
computer with one core and using the compiler Intel-ifort version
$12.1.0$.
To obtain this level of speed, it is important to notice that there are two contributions dominating 
in two different phase space regions associated with the two decay modes of the $W$ bosons, namely 
$W^\pm \to l^{\pm}\overset{\textbf{\fontsize{0.5pt}{0.5pt}\selectfont(---)}}{\nu_{l}}$ and 
$W^\pm \to l^{\pm}\overset{\textbf{\fontsize{0.5pt}{0.5pt}\selectfont(---)}}{\nu_{l}} \gamma$. 
This means that there are two different positions of the on-shell $W$ pole in
the phase space. For efficient Monte
Carlo generation, we divide the phase space into two separate regions 
to consider these two possibilities and then sum the contributions to get the
total result. The regions are generated as double EW boson production
as well as $W$ production with (approximately) on-shell $W^+\to
\ell^+\nu_l\gamma$ (or  $W^-\to \ell^-\bar\nu_l\gamma$) three-body decay,
respectively, and
are chosen according to whether $m(\ell^+\nu_l\gamma)$ or $m(\ell^+\nu_l)$ is closer to $M_W$. 


\section{Numerical results}
\label{sec:results}
In this section, we present results for the integrated cross section and
for various differential distributions. As EW input parameters, we use  $M_W=80.385 \GeV$, $M_Z=91.1876 \GeV$ and
$G_F=1.16637\times 10^{-5}\GeV^{-2}$. We then use tree-level relations
to calculate the weak mixing angle and the electromagnetic coupling from
these.  As parton distribution functions we use the MSTW2008 parton
distribution functions~\cite{Martin:2009iq} with
$\alpha_s^\text{LO}(M_Z)=0.13939$ and
$\alpha_s^\text{NLO}(M_Z)=0.12018$. 
The $W$ decay width is calculated as $\Gamma_W = 2.09761 \GeV$.  
With the lepton-photon separation $R_{l\gamma} > 0.4$ (see below), we can set the charged lepton masses 
to zero because they are very small compared to the minimum invariant mass of the lepton-photon system, 
which is about $10 \GeV$. We work in the five-flavor scheme and use the 
$\overline{MS}$ renormalization of the strong coupling constant with the top quark decoupled from the running of $\alpha_s$. 
However, the top-loop contribution is
explicitly included in the virtual amplitudes, using $m_t=173.1\GeV$. To have a large phase space for
QCD radiation, we choose inclusive cuts defined as 
\begin{align}
  p_{T(j,l)} &> 20 \GeV&  p_{T(\gamma)} &> 30 \GeV&  \slashed p_T &> 30 \GeV& \crn 
  |y_j| &< 4.5&   |y_{l}| &< 2.5&   |y_{\gamma}| &< 2.5& \crn
  R_{jl} &> 0.4&   R_{l\gamma} &> 0.4&   R_{j\gamma} &> 0.7,&
\end{align}
where the missing energy is associated with the neutrino. The anti-$k_t$
algorithm~\cite{Cacciari:2008gp} with a cone radius of $R=0.4$ is used
to cluster partons into jets. 
To deal with the real photon in the final state, we use the smooth isolation cut 
\`a la Frixione~\cite{Frixione:1998jh}. Events are accepted if 
\bea
\sum_{i\in \text{partons}} p_{T,i}\theta(R-R_{\gamma i}) \le p_{T,\gamma}\frac{1-\cos R}{1-\cos\delta_0} \quad \forall R<\delta_0
\label{eq:Frixione cut}
\eea
with $\delta_0 = 0.7$. 
As dynamical factorization and renormalization scale, we use as central
value 
\bea \mu_{0}=a\left(\sum_{\text{jets}} p_{T,i} \text{e}^{b|y_{i}-y_{12}|} + 
p_{T,\gamma} + E_{T,W}\right),
\label{eq:define_mu0}
\eea 
where $E_{T,W}=(p^2_{T,W} + m_W^2)^{1/2}$, with $m_W$ being the reconstructed mass, denotes the 
transverse energy of the W boson and $y_{12} = (y_1 + y_2)/2$ the average rapidity of the two hardest jets. 
The parameters $a$ and $b$ are arbitrary and we choose $a=1/2$ and $b=1$ such that the 
first term in the right hand side of \eq{eq:define_mu0} is equal to the invariant mass, $m_{jj}$, of the two hardest jets in the large $|y_1 - y_2|$ limit and for $p_{T,j_1} \approx p_{T,j2}$. 
If $p_{T,j_1} \gg p_{T,j2}$ then it is much larger than $m_{jj}$. 
For small $\Delta y_\text{tags}$ this contribution approaches $\sum_{\text{jets}} p_{T,i}/2$. 
It was suggested first in \bib{Ellis:1992en} 
in the framework of di-jet production and was proved to be appropiate for
$W^+W^+jj$ production in Ref.~\cite{Campanario:2013gea}.

\begin{figure}[t!]
  \centering
  \includegraphics[width=0.83\columnwidth]{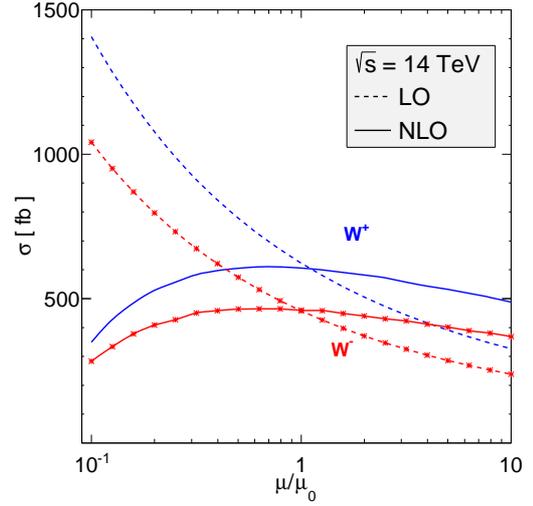}
\caption{Scale dependence of the LO and NLO cross sections at the LHC. 
The curves with and without stars are for $W^-\gamma jj$ and $W^+\gamma jj$ productions, 
respectively. The reference scale $\mu_0$ is defined in \eq{eq:define_mu0} and the text.}
\label{fig:scale}
\end{figure}

At $\mu_o$ with this set up we obtain 
$\sigma_{LO} = 622.7 \, \pm \, 0.1 \fb$ 
$( 457.6 \, \pm \,0.1 \fb )$
and 
$\sigma_{NLO} =  605.0 \, \pm \, 0.3 \fb \,
(459.6 \, \pm \, 1.2 \fb )$ for 
$W^{+}\gamma jj$ $(W^{-}\gamma jj)$ production, with the $W$ decaying into the first generation of leptons.
Here the errors are the Monte Carlo errors of the calculation.
The \kfac defined as $  K\equiv \sigma_{NLO}/\sigma_{LO}$, is
$0.97 \, (1.00)$. 
\begin{figure*}[ht!]
  \centering
  \includegraphics[width=0.83\columnwidth]{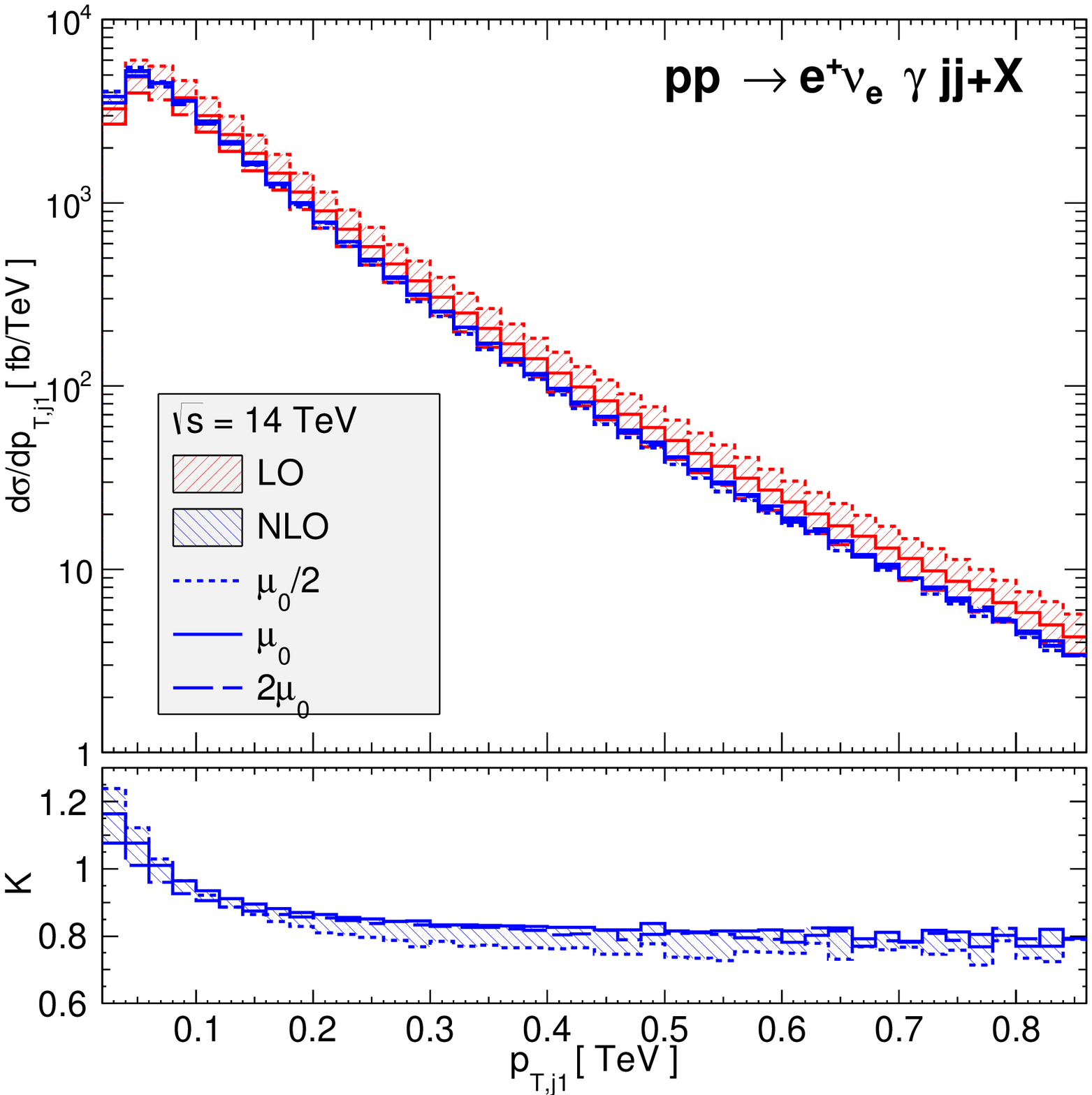}
  \includegraphics[width=0.83\columnwidth]{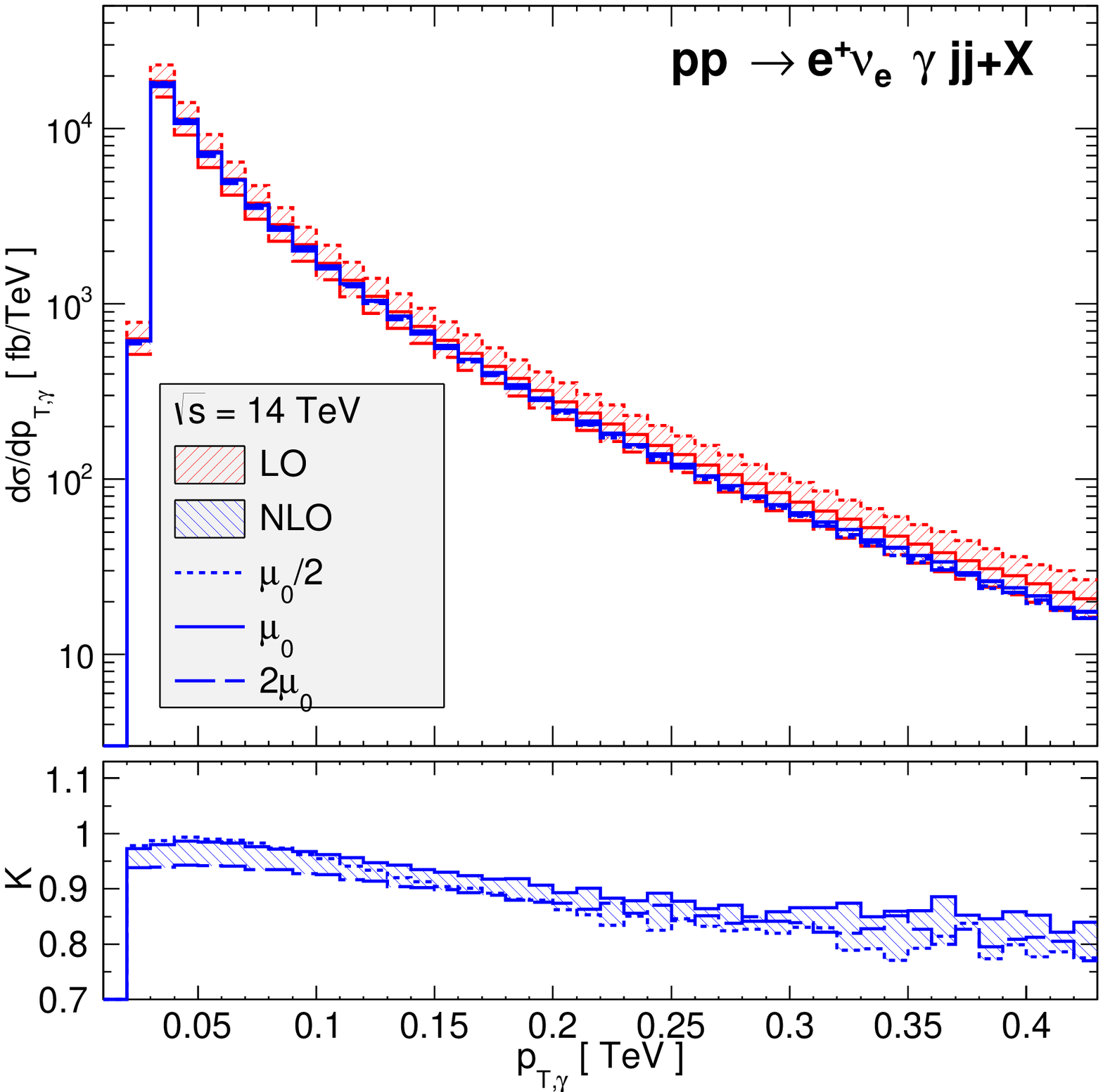}
  \includegraphics[width=0.83\columnwidth]{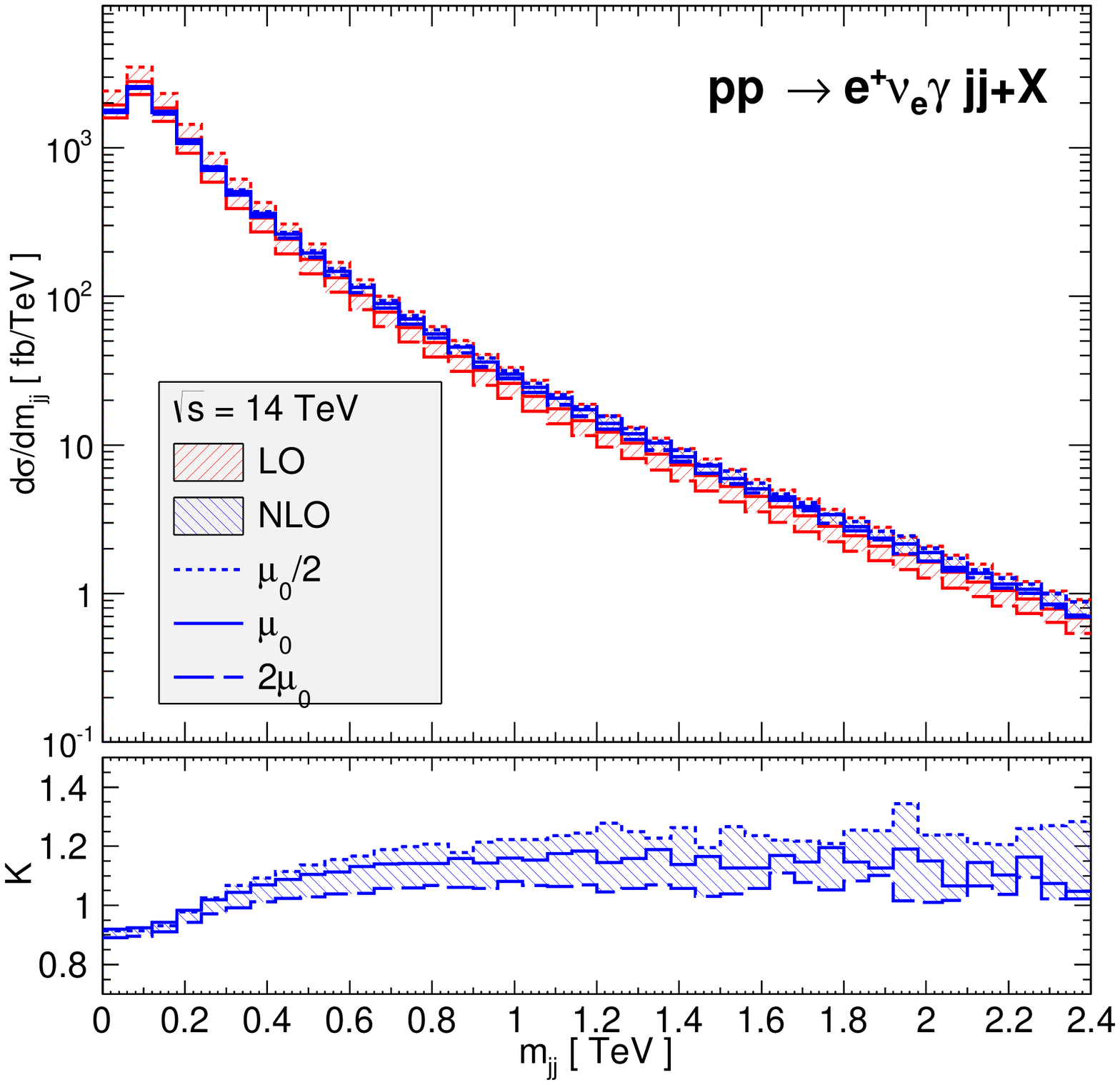}
  \includegraphics[width=0.83\columnwidth]{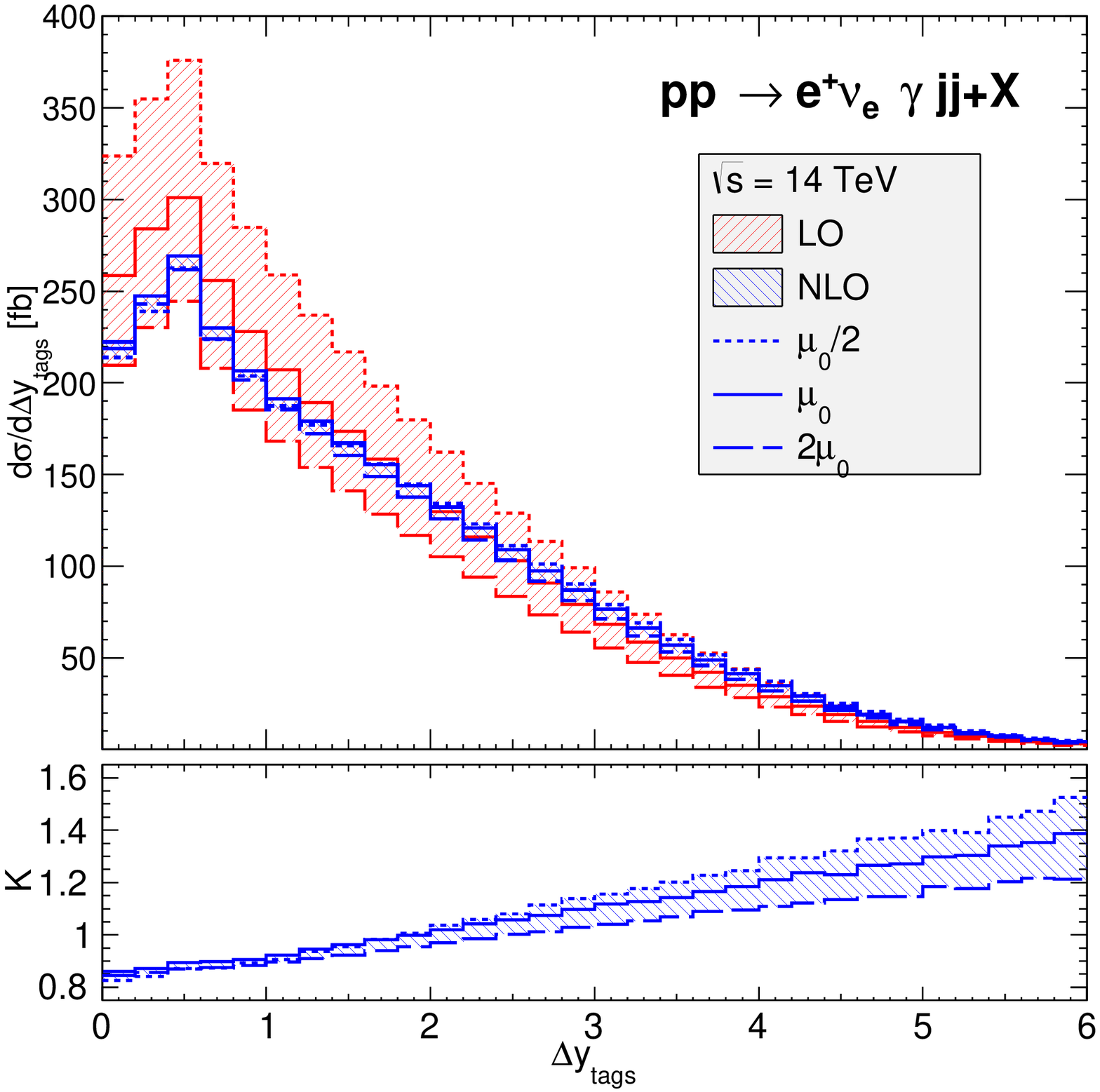}
  \caption{Differential cross sections, for QCD-induced $W\gamma jj$ production at LO and NLO, 
with inclusive cuts are shown for the transverse momenta of the hardest jet (top
left) and the photon (top right), the invariant mass  (bottom left) of the two tagging jets ordered by $p_T$. 
The distributions of the rapidity separation between the two jets are in the bottom right panel. 
The bands describe $\mu_0/2 \le \mu_F=\mu_R\le 2\mu_0$ variations. 
The \kfac bands are due to the scale variations of the NLO results, 
with respect to $\sigma_\text{LO}(\mu_0)$. 
The dots in the small panels are for the central scale, while the two solid lines correspond to $\mu_F = \mu_R=2\mu_0$ and $\mu_0/2$.
}
\label{dist_NLO_jets_inc}
\end{figure*}
The result depends on the factorization and renormalization scales
since we only calculate at fixed order in perturbative
QCD. Fig.~\ref{fig:scale} shows, both for $W^+ \gamma jj $ and $W^-
\gamma jj$ production, that the dependence of the cross section on
the factorization and renormalization scale, which are set equal for
simplicity, is significantly reduced when calculating the NLO QCD
corrections. If we vary the two scales separately, a small dependence on
$\mu_F$ is observed, while 
the $\mu_R$ dependence is similar to the behavior shown in
\fig{fig:scale}. 

\begin{figure*}[t!]
  \centering
  \includegraphics[width=0.95\columnwidth]{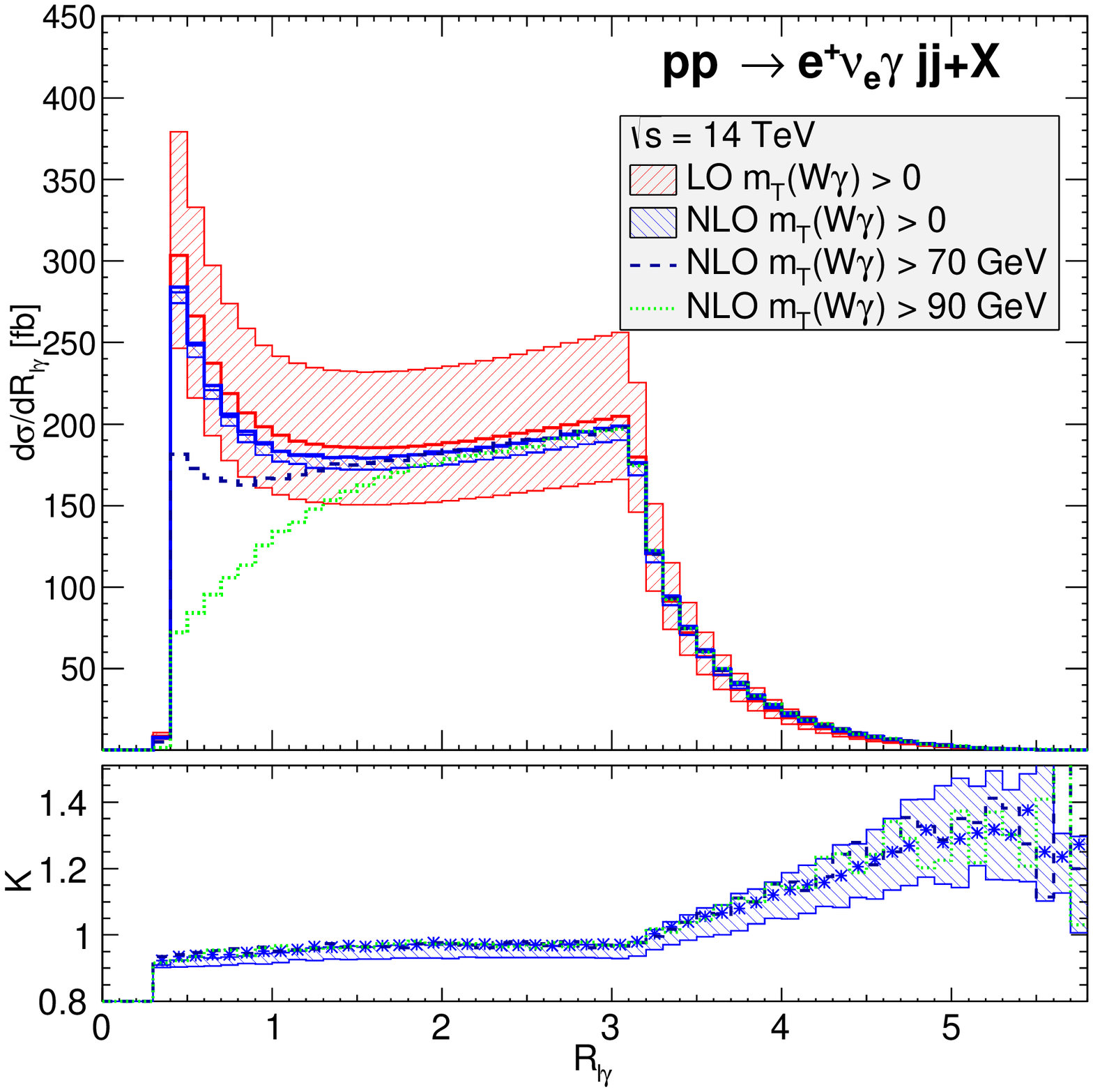}
  \includegraphics[width=0.95\columnwidth]{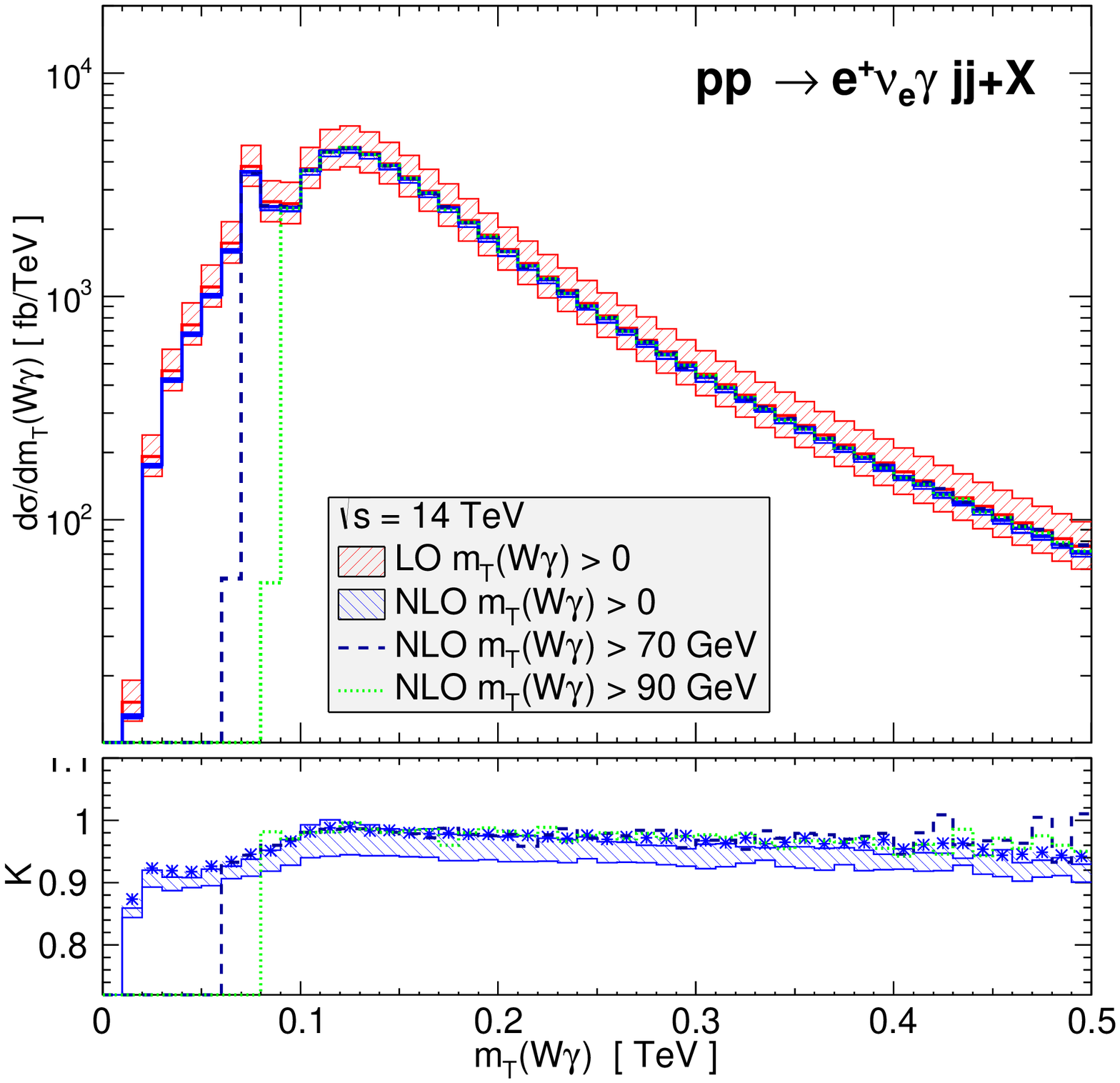}
  \includegraphics[width=0.95\columnwidth]{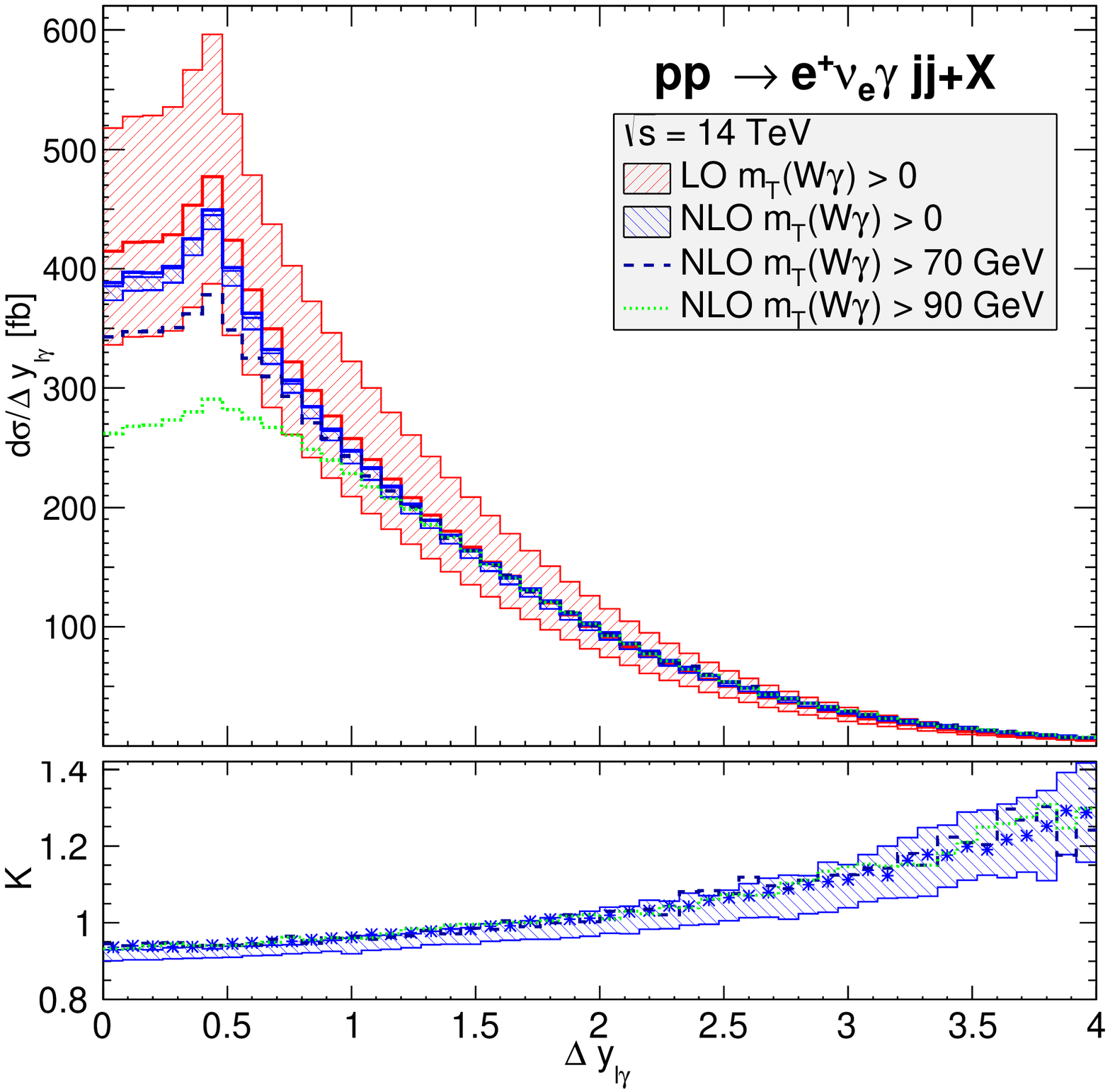}
  \includegraphics[width=0.95\columnwidth]{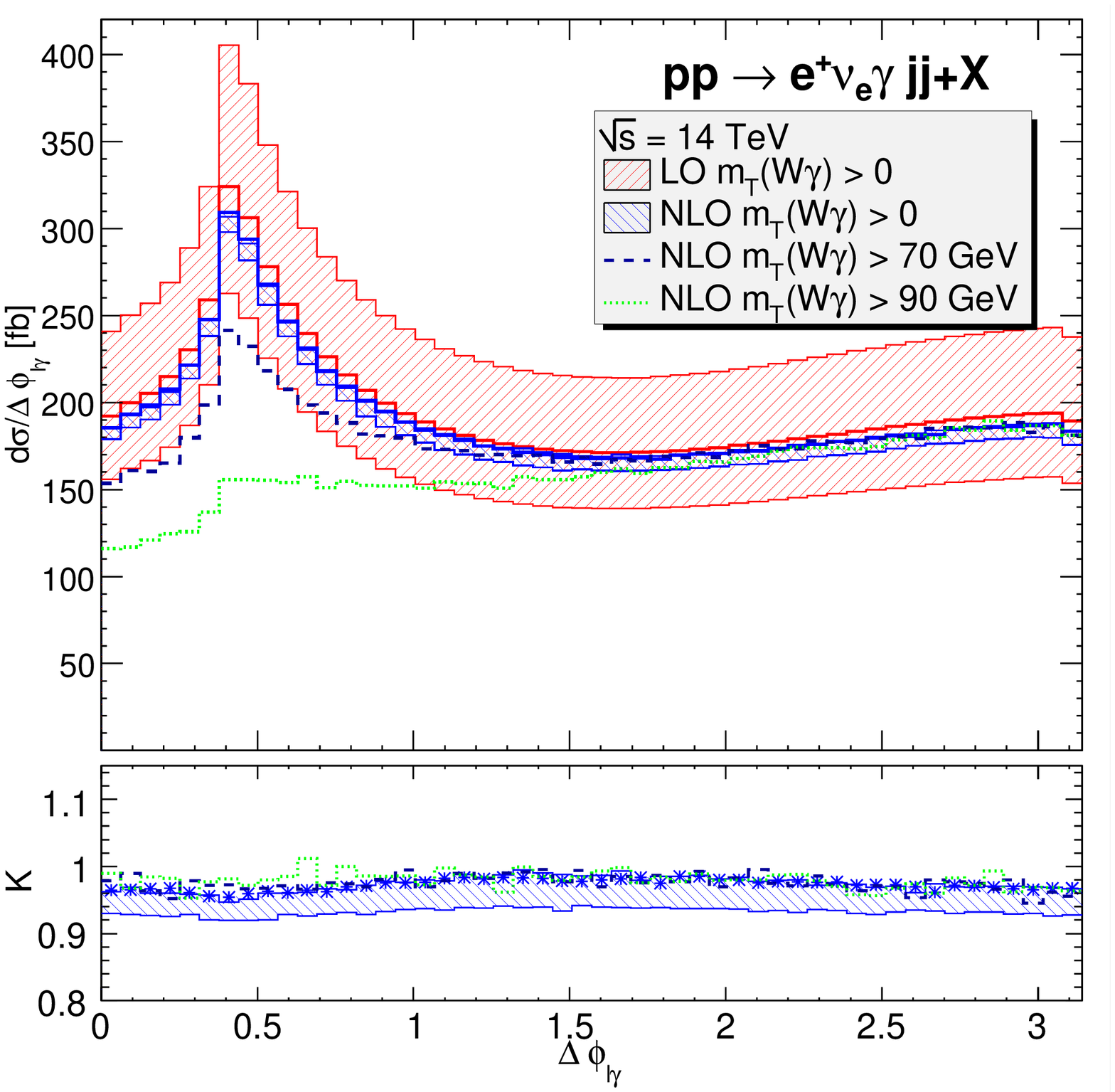}
  \caption{Differential cross sections, for the QCD-induced channels at LO and NLO, 
with inclusive cuts and for different
values of the $m_{T,(W\gamma)}$ cut. In the upper row the distributions $R_{l\gamma}$ (left) and the transeverse cluster
energy of the $W\gamma$ system $m_{T,(W\gamma)}$ (right) are shown.
The lower row shows the rapidity (left) and azimuthal angle (right) separation of the photon and lepton.
The bands on the distributions describe $\mu_0/2 \le \mu_F=\mu_R\le 2\mu_0$ variations for $m_{T,(W\gamma)}>0$. 
The corresponding \kfac bands are due to the scale variations of the NLO results, 
with respect to $\sigma_\text{LO}(\mu_0)$. 
The curves with stars in the narrow panels below the distributions are for the central scale for $m_{T,(W\gamma)}>0$. 
The $K$-factors at $\mu_0$ for the other cuts are also shown. 
}
\label{fig:finalrad}
\end{figure*}
In the following, distributions for the 
$W^+ \gamma j j$ production channel will be
presented. The results for $W^- \gamma jj$ production are similar. Fig.~\ref{dist_NLO_jets_inc} shows in the top row the
differential LO and NLO cross sections of the transverse momentum of the
hardest jet (left) and the photon (right), and in the lower row, the
invariant mass (left) and rapidity separation (right) of the two tagging jets ordered by $p_T$. 
To give a measure of scale uncertainty, we also include with bands the results for
$\mu_{F}=\mu_{R}=\mu= 2^{\pm 1} \mu_0$. The small panels show the
differential \kfac{}s, defined as the ratio of the NLO to the LO results.

The differential distributions are less sensitive at NLO to the scale
variation than at LO and the relative scale uncertainty is equally
distributed in the entire $p_{T,j_1}$ and $p_{T,\gamma}$ spectrum. The
phase space shows a non-trivial dependence with \kfac{}s varying, for
$\mu = \mu_0$, from $1.2$ to $0.8$ for the $p_{T}$ distribution of the
hardest jet and from $0.95$ to $0.8$ for the transverse momenta of the
photon in the ranges shown.

In the bottom panels, we observe a similar significant reduction of the
scale uncertainties for the $m_{jj}$ (left) and the  $\Delta y_\text{tags}$
(right) differential distributions, with the \kfac of the invariant 
mass distributions varying from about 0.9 to
1.2 at 2.4 TeV and with a fairly constant slope and the \kfac for the
rapidity difference of the two leading tagging jets varying from 0.85 to 1.4
in the range showed.

Finally in Fig.~\ref{fig:finalrad}, we plot in the left the differential distribution
of the separation in the rapidity azimuthal-angle plane of the lepton
and photon, $R_{l\gamma}$, and  on the the right the transverse cluster
mass of the $W\gamma$ system defined as (see e.g. \bib{Baur:1993ir})
\bea
 m_{T,W\gamma} = \left(\left[(m_{l\gamma}^2 +
      p_{T_{l\gamma}}^{2})^{\frac{1}{2}} + \displaystyle{\not}p_{T} \right]^2 -
    (  {\boldsymbol p_{\boldsymbol T} }_{l\gamma} + { \displaystyle{\not}
      \boldsymbol p_{\boldsymbol T} } )^{2} \right)^{\frac{1}{2}}. \nonumber\\[-3mm]
\eea
In those plots, one can observe how the photon radiated off the lepton can be effectively removed by imposing a cut on the
transverse cluster
mass. This radiative W decay represents a simple QED process (bottom left
diagram of Fig.~\ref{fig:feynTree}),
which diminishes the sensitivity to anomalous couplings, which might enter in e.g. the top left
diagram of Fig.~\ref{fig:feynTree}. For
$m_{T,W\gamma}>90 \GeV$, the radiative decay peak at
$m_{T,W\gamma}=m_{W}$ is eliminated, affecting
mainly the region of small $R_{l\gamma}$~(left). Furthermore,
the NLO cross section is reduced by approximately 15\% showing
the efficiency of the cut. 

The $R_{l\gamma}$ distribution in Fig.~\ref{fig:finalrad} shows a 
sudden increase of the \kfac starting at $\pi$, which correlates to the sudden fall of 
the differential cross section. This discontinuity in the slope can be explained as follows. 
The $R$ separation is defined as $R_{l\gamma} = [(\Delta y_{l\gamma})^2 + (\Delta\phi_{l\gamma})^2]^{1/2}$ 
where $\Delta\phi_{l\gamma} \in [0,\pi]$. For $0 < R_{l\gamma} < \pi$, the dominant contribution 
comes from the $\Delta y_{l\gamma} \approx 0$ region (see the $\Delta y_{l\gamma}$ distribution in Fig.~\ref{fig:finalrad}), and the behavior of the \kfac is given by the one of the 
$\Delta \phi_{l\gamma}$ distribution also displayed in Fig.~\ref{fig:finalrad}, which is rather flat. 
For $R_{l\gamma} > \pi$, the rapidity separation must increase and the \kfac is similar to the one 
of the $\Delta y_{l\gamma}$ distribution. 

The above results for various differential distributions show that our default scale choice defined in 
\eq{eq:define_mu0} and the text can make the LO results quite similar to the NLO ones, 
with the difference being smaller than $20\%$ in most cases. The exceptional cases are the distributions of 
$\Delta y_\text{tags}$ (see \fig{dist_NLO_jets_inc}) and $\Delta y_{l\gamma}$ (see \fig{fig:finalrad}). 
Here we observe that the \kfac increases with large rapidity separations. This indicates that the default scale choice is 
too large at large rapidity separations, making the LO results too small. We have tried a different scale choice, 
using \eq{eq:define_mu0} with $a=1/2$ and $b=0$, and found that the NLO results, for the distributions shown, agree with 
the ones obtained with the default scale within $10\%$, while the two scale choices at LO produce differences as large as a factor of 2 for the $m_{jj}$ and $\Delta y_\text{tags}$ distributions.
We also found that the new scale choice makes the \kfac{}s decrease well below one with increasing invariant mass or rapidity separation of the two 
hardest jets. 

\section{Conclusions}
\label{sec:con}
In this paper, we have reported first results for $W^\pm \gamma jj + X$ production at order $\order{\alpha_s^3 \alpha^3}$,
including the leptonic decays, full off-shell and finite width effects as
well as all spin correlations. The NLO QCD corrections to the total cross section are small but they
exhibit non-trivial phase space dependencies, reaching up to $40\%$, and lead to shape changes of the distributions. Hence, they should be taken
into account for precise measurements at the LHC.

Our code will be publicly available as part of the {\texttt{VBFNLO}} 
program~\cite{Arnold:2008rz,Arnold:2012xn}, thereby further studies of the QCD corrections with different kinematic cuts can be easily done.

\begin{acknowledgements}
We acknowledge the support from the Deutsche Forschungsgemeinschaft
via the Sonderforschungsbereich/Transregio SFB/TR-9 Computational
Particle Physics.  FC is funded by a Marie Curie fellowship
(PIEF-GA-2011-298960) and partially by MINECO (FPA2011-23596) and by
LHCPhenonet (PITN-GA-2010-264564).  MK is supported by the
Graduiertenkolleg 1694 ``Elementarteilchenphysik bei h\"ochster
Energie und h\"ochster Pr\"azision''.
\end{acknowledgements}

\appendix
\section{Results at one phase-space point}
\label{appendixA}
In this appendix, we provide results at a random phase-space point to
facilitate comparisons with our results. We focus on the virtual
amplitudes of the five benchmark subprocesses \eq{eq:subproc}.
The amplitudes of all other subprocesses can be obtained via crossing.
The phase-space point for the process $j_1 j_2 \to j_3 j_4 e^+ \nu_e
\gamma$ is given in \tab{table_PSP_2to5}.
\begin{table*}[th]
 \begin{footnotesize}
 \begin{center}
 \caption{\label{table_PSP_2to5}{Momenta (in GeV) at a random phase-space point for $j_1 j_2 \to j_3 j_4 e^+ \nu_e \gamma$ subprocesses.}}
\begin{tabular}{l | r@{.}l r@{.}l r@{.}l r@{.}l}
& \multicolumn{2}{c}{ $E$}
& \multicolumn{2}{c}{ $p_x$}
& \multicolumn{2}{c}{ $p_y$}
& \multicolumn{2}{c}{ $p_z$}
\\
\hline
$j_1$  & 32&0772251055223 & 0&0  & 0&0  &  32&0772251055223  \\
$j_2$  & 2801&69305619768 & 0&0  & 0&0  &  -2801&69305619768  \\
$j_3$  & 226&525314156010 & -10&2177083492279 & -1&251308382450315$\times 10^{-15}$ & -226&294755550298 \\
$j_4$  & 327&281588297290 & -6&48554750244653 & -10&1061447270513 & -327&061219882068 \\
$e^+$  & 646&824307052136 & 36&0746355875450 & -26&0379256562231 & -645&292438579767 \\
$\nu_e$  & 1598&85193997112 & -2&88431497177613 & 24&4490976584709 & -1598&66239347157 \\
$\gamma$  & 34&2871318266438 & -16&4870647640944 & 11&6949727248035 & 27&6949763915464\\
\hline
\end{tabular}\end{center}
 \end{footnotesize}
\end{table*}
\begin{table*}[th]
 \begin{footnotesize}
 \begin{center}
\caption{\label{table_PSP_QCD_uduu}{QCD interference amplitudes $2\text{Re}(\mathcal{A}_\text{NLO}\mathcal{A}^{*}_\text{LO})$
for $u \bar{d} \to \bar{u} u e^+ \nu_e \gamma$ subprocess.}}
\begin{tabular}{l | r@{.}l r@{.}l r@{.}l}
& \multicolumn{2}{c}{ $1/\epsilon^2$}
& \multicolumn{2}{c}{ $1/\epsilon$}
& \multicolumn{2}{c}{ finite}
\\
\hline
I operator  & 208&754750693775 & 346&823893959906 & 214&565536875302 \\
loop  & -208&754750694041 & -346&823893964206 & 1309&48703231438 \\
I+loop  & -2&661124653968727$\times 10^{-10}$ & -4&300034106563544$\times 10^{-9}$ & 1524&05256918968 \\
\hline
\end{tabular}\end{center}
 \end{footnotesize}
\end{table*}
\begin{table*}[th!]
 \begin{footnotesize}
 \begin{center}
 \caption{\label{table_PSP_QCD_udcc}{QCD interference amplitudes $2\text{Re}(\mathcal{A}_\text{NLO}\mathcal{A}^{*}_\text{LO})$
for $u \bar{d} \to \bar{c} c e^+ \nu_e \gamma$ subprocess.}}
\begin{tabular}{l | r@{.}l r@{.}l r@{.}l}
& \multicolumn{2}{c}{ $1/\epsilon^2$}
& \multicolumn{2}{c}{ $1/\epsilon$}
& \multicolumn{2}{c}{ finite}
\\
\hline
I operator  & 204&162116147897 & 338&462143954872 & 193&947096152034 \\
loop  & -204&162116148143 & -338&462143959044 & 1250&53101019255 \\
I+loop  & -2&459898951201467$\times 10^{-10}$ & -4&172022727289004$\times 10^{-9}$ & 1444&47810634458 \\
\hline
\end{tabular}\end{center}
 \end{footnotesize}
\end{table*}
\begin{table*}[th]
 \begin{footnotesize}
 \begin{center}
\caption{\label{table_PSP_QCD_uddd}{QCD interference amplitudes $2\text{Re}(\mathcal{A}_\text{NLO}\mathcal{A}^{*}_\text{LO})$
for $u \bar{d} \to \bar{d} d e^+ \nu_e \gamma$ subprocess.}}
\begin{tabular}{l | r@{.}l r@{.}l r@{.}l}
& \multicolumn{2}{c}{ $1/\epsilon^2$}
& \multicolumn{2}{c}{ $1/\epsilon$}
& \multicolumn{2}{c}{ finite}
\\
\hline
I operator  & 211&035586302262 & 351&015498463760 & 217&505955345503 \\
loop  & -211&035586301469 & -351&015498460714 & 1288&70122715328 \\
I+loop  & 7&927951628516894$\times 10^{-10}$ & 3&046068286494119$\times 10^{-9}$ & 1506&20718249878 \\
\hline
\end{tabular}\end{center}
 \end{footnotesize}
\end{table*}
\begin{table*}[th]
 \begin{footnotesize}
 \begin{center}
\caption{\label{table_PSP_QCD_udss}{QCD interference amplitudes $2\text{Re}(\mathcal{A}_\text{NLO}\mathcal{A}^{*}_\text{LO})$
for $u \bar{d} \to \bar{s} s e^+ \nu_e \gamma$ subprocess.}}
\begin{tabular}{l | r@{.}l r@{.}l r@{.}l}
& \multicolumn{2}{c}{ $1/\epsilon^2$}
& \multicolumn{2}{c}{ $1/\epsilon$}
& \multicolumn{2}{c}{ finite}
\\
\hline
I operator  & 204&420606439876 & 338&890671930900 & 194&192653175338 \\
loop  & -204&420606439076 & -338&890671927825 & 1255&72258287559 \\
I+loop  & 8&000995421753032$\times 10^{-10}$ & 3&075001586694270$\times 10^{-9}$ & 1449&91523605093 \\
\hline
\end{tabular}\end{center}
 \end{footnotesize}
\end{table*}
\begin{table*}[th]
 \begin{footnotesize}
 \begin{center}
\caption{\label{table_PSP_QCD_ggud}{QCD interference amplitudes $2\text{Re}(\mathcal{A}_\text{NLO}\mathcal{A}^{*}_\text{LO})$
for $g g \to \bar{u} d e^+ \nu_e \gamma$ subprocess.}}
\begin{tabular}{l | r@{.}l r@{.}l r@{.}l}
& \multicolumn{2}{c}{ $1/\epsilon^2$}
& \multicolumn{2}{c}{ $1/\epsilon$}
& \multicolumn{2}{c}{ finite}
\\
\hline
I operator  & 0&134340391976220 & 1&216201313632209$\times 10^{-2}$  & 0&200038523086631 \\
loop  & -0&134340391970929 & -1&216201313220419$\times 10^{-2}$ & 0&164526176760187 \\
I+loop  & 5&291489468817190$\times 10^{-12}$ & 4&117898730338077$\times 10^{-12}$ & 0&364564699846818 \\
\hline
\end{tabular}\end{center}
 \end{footnotesize}
\end{table*}
In the following we provide the squared amplitude averaged over the initial-state 
helicities and colors. We also set $\alpha = \alpha_s = 1$ for simplicity. 
The top quark is decoupled from the running of $\alpha_s$. However, its contribution 
is explicitly included in the one-loop amplitudes.
At tree level, we have
\begin{align}
  \overline{|\mathcal{A}_\text{LO}^{u\bar{d}\rightarrow \bar{u}u}|}^2 &= 245.933396692488 ,\nonumber \\
  \overline{|\mathcal{A}_\text{LO}^{u\bar{d}\rightarrow \bar{c}c}|}^2 &= 240.522826586251 ,\nonumber \\
  \overline{|\mathcal{A}_\text{LO}^{u\bar{d}\rightarrow \bar{d}d}|}^2 &= 248.620442839372 ,\nonumber \\
  \overline{|\mathcal{A}_\text{LO}^{u\bar{d}\rightarrow \bar{s}s}|}^2 &= 240.827353287120 ,\nonumber \\
  \overline{|\mathcal{A}_\text{LO}^{gg\rightarrow \bar{u}d}|}^2 &= 9.739448965670859\times 10^{-2}.
\end{align}
The interference amplitudes
$2\text{Re}(\mathcal{A}_\text{NLO}\mathcal{A}^{*}_\text{LO})$,
for the one-loop corrections (including counterterms) and the I-operator contribution as defined in \bib{Catani:1996vz},
are given in \tab{table_PSP_QCD_uduu}, \tab{table_PSP_QCD_udcc}, \tab{table_PSP_QCD_uddd}, \tab{table_PSP_QCD_udss} and \tab{table_PSP_QCD_ggud}. Here we use the following
convention for the one-loop integrals, with $D=4-2\epsilon$,
\bea
T_0 = \frac{\mu_R^{2\epsilon}\Gamma(1-\epsilon)}{i\pi^{2-\epsilon}}\int d^D q \frac{1}{(q^2 - m_1^2 + i0)\cdots}.
\eea
This amounts to dropping a factor ${(4\pi)^\epsilon}/{\Gamma(1-\epsilon)}$
both in the virtual corrections and the I-operator.
Moreover, the conventional dimensional-regularization method~\cite{'tHooft:1972fi}
with $\mu_{R} = M_Z$ is used. 
Changing from the conventional dimensional-regularization method to the dimensional reduction scheme 
induces a finite shift. 
This shift can be easily found by observing that 
the sum $|\mathcal{A}_\text{LO}|^2+2\text{Re}(\mathcal{A}_\text{NLO}\mathcal{A}^{*}_\text{LO})$ 
must be unchanged as explained in \bib{Catani:1996pk}. 
Thus, the shift on\\ 
$2\text{Re}(\mathcal{A}_\text{NLO}\mathcal{A}^{*}_\text{LO})$ is opposite to the shift 
on the Born amplitude squared, which in turn is given by the following change in the strong coupling constant, 
see e.g. \bib{Kunszt:1993sd},
\bea
\alpha_s^{\overline{DR}} = \alpha_s^{\overline{MS}}\left(1+\frac{\alpha_s}{4\pi}\right).
\eea
The shift on the I-operator contribution can easily be calculated using the rule given in \bib{Catani:1996vz}. 

\bibliographystyle{spphys}
\bibliography{QCDVVjj}

\end{document}